# Investigating operation of the Internet in orbit: Five years of collaboration around CLEO

L. Wood,[@] W. Ivancic,[+] W. Eddy,[$+] D. Stewart,[$+] J. Northam,[*] C. Jackson[*]
[@]Cisco Systems Global Government Solutions Group, Bedfont Lakes, London, UK.
[*]Surrey Satellite Technology Ltd, University of Surrey, Guildford, UK.
[+]NASA Glenn Research Center, Cleveland, Ohio.
[$]Verizon Federal Network Systems, Cleveland, Ohio.
Correspondence to: **lwood@cisco.com**   http://info.ee.surrey.ac.uk/Personal/L.Wood/cleo/

## Introduction

The Cisco router in Low Earth Orbit (CLEO) was launched into space as an experimental secondary payload onboard the UK Disaster Monitoring Constellation (UK-DMC) satellite in September 2003. The UK-DMC satellite is one of an increasing number of DMC satellites in orbit that rely on the Internet Protocol (IP) for command and control and for delivery of data from payloads. The DMC satellites, built by Surrey Satellite Technology Ltd (SSTL), have imaged the effects of Hurricane Katrina, the Indian Ocean Tsunami, and other events for disaster relief under the International Space and Major Disasters Charter.

It was possible to integrate the Cisco mobile access router into the UK-DMC satellite as a result of the DMC satellites' adoption of existing commercial networking standards, using IP over Frame Relay over standard High-Level Data Link Control, or HDLC (ISO 13239) on standard serial interfaces. This approach came from work onboard SSTL's earlier UoSAT-12 satellite [1].

## First tests of CLEO

A large team came together in June 2004 to show that a commercial Internet router could operate in a space vacuum environment. With SSTL and Cisco was NASA Glenn Research Center, who had previously worked with Cisco Systems under a US Space Act Agreement on mobile networking. Also involved for this first demonstration were a number of US military organisations, testing IP-based Virtual Mission Operations Center (VMOC) software from General Dynamics. The VMOC successfully commanded the CLEO router and tasked the UK-DMC satellite from a field tent at Vandenberg Air Force Base, using the Internet to SSTL's ground station [2]. CLEO's use of Mobile IP and mobile networking were also evaluated.

The CLEO ground-based testbed (figure 1), shipped to NASA Glenn, was key to this successful demonstration. That enabled NASA Glenn to become familiar with DMC satellite operation and to provide working router configurations that could be uploaded during very limited operational time with the on-orbit router, when the satellite was not carrying out its primary imaging mission. Direct access to the CLEO router on orbit was later demonstrated at several conferences.

## First IPv6 from space

The NASA/SSTL/Cisco team then upgraded SSTL's network infrastructure to support IP version 6, and again used their testbed to become the first to successfully test IPv6 in space with CLEO in March 2007 [3]. CLEO was also configured for IPv4 IPSec use at that time.

## First delay-tolerant network bundles in space

Use of the ground-based testbed then shifted away from simply configuring CLEO to programming the SSTL imaging computers that sit alongside CLEO. NASA Glenn is now able to make changes to the RTEMS operating system code on the SSTL computer in its testbed, test those changes, and pass code to SSTL to be uploaded and tested in orbit.

Our team worked to improve *Saratoga*, the fast UDP/IP-based transfer protocol that was developed at SSTL in 2004 to replace an implementation of CCSDS CFDP that was considered slow. *Saratoga* is capable of fully utilizing a link and handles very asymmetric link environments. This has been described to the IETF [4].

In passing periodically over ground stations with limited contact times, the DMC satellites experience link disruption and intermittent connectivity. This is a form of 'Delay-Tolerant Networking' (DTN) [5]. The 'Bundle Protocol' developed by the IRTF DTN research group (DTNRG) is one approach to handling DTN [6].

Our team became the first to successfully demonstrate use of the DTNRG's Bundle Protocol from space, by adopting *Saratoga* as a bundle convergence layer [7] to download an Earth image (figure 2). This 150-megabyte image was downloaded across multiple passes, using proactive bundle fragmentation to avoid the effects of link disruption [8].

Although bundling over the Internet Protocol and *Saratoga* adds no significant benefits to SSTL's existing operational model, the practical experience gained has

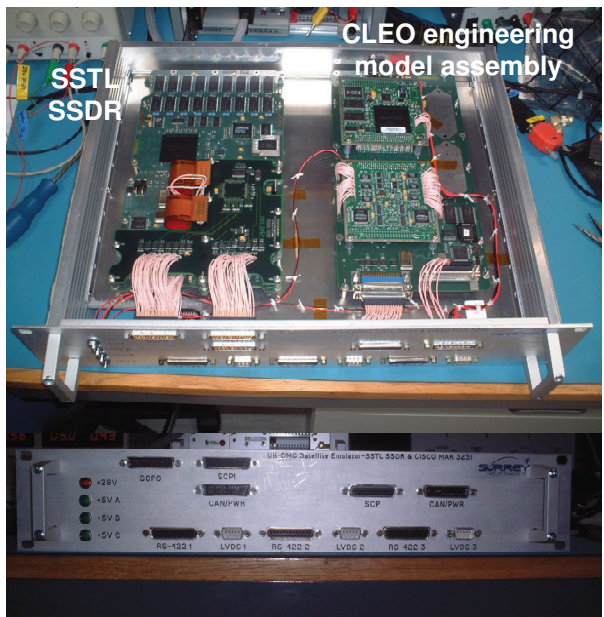

a. top view, before adding fans and heatsinks.
b. front view of ports

**Figure 1 – CLEO ground-based testbed**

helped our team in identifying issues with the Bundle Protocol's design that must be addressed.

The design of the current Bundle Protocol ignores the *end-to-end principle*. The Bundle design does not include error detection or rejection for ensuring reliability in headers or payload data. Adding reliability into the existing Bundle design, by leveraging the not-yet-complete security protocols proposed by the DTN research group, is possible as a workaround, but there are also drawbacks to taking this approach [9].

## Summary

The DMC satellites and their use of the Internet Protocol for imaging transfers provide working operational examples of mission-critical use of IP for sensor networks, allowing cost-effective development and easy integration with the terrestrial Internet for data delivery. Use of the efficient *Saratoga* transfer protocol and IP to carry sensor data operates well every day.

The CLEO router in orbit, made possible by the DMC satellites' use of IP, has now been in space for over five years and has been used in orbit over one hundred times.

The team of NASA Glenn Research Center, Cisco Systems and Surrey Satellite Technology Ltd has now gained considerable practical experience and leadership both with taking the Internet into space, and with delay-tolerant networking. The success of these early experiments has made follow-on work possible. That work includes taking Internet routers to geostationary orbit onboard commercial communication satellites.

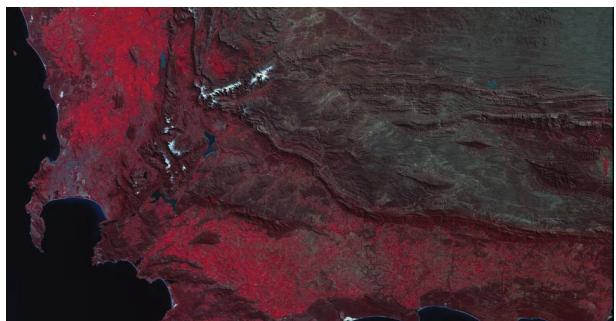

**Figure 2 – Image delivered via the Bundle Protocol Cape of Good Hope and Karoo desert**

We thank everyone who has been involved in and who has contributed to this work thus far.

## References

[1] K. Hogie, E. Criscuolo and R. Parise, "Using standard Internet Protocols and applications in space," Computer Networks, vol. 47 no. 5, pp. 603-650, April 2005.

[2] L. Wood, W. Ivancic, D. Hodgson, E. Miller, B. Conner, S. Lynch, C. Jackson, A. da Silva Curiel, D. Shell, J. Walke and D. Stewart, "Using Internet nodes and routers onboard satellites," International Journal of Satellite Communications and Networking, vol. 25 no. 2, pp. 195-216, March/April 2007.

[3] W. Ivancic, D. Stewart *et al*., "IPv6 and IPSec Tests of a Space-Based Asset, the Cisco router in Low Earth Orbit (CLEO)," NASA Technical Memorandum 2008-215203, May 2008.

[4] L. Wood, J. McKim, W. M. Eddy, W. Ivancic and C. Jackson, "Saratoga: A Scalable File Transfer Protocol," work in progress as an internet-draft, draft-wood-tsvwg-saratoga, October 2008.

[5] V. Cerf *et al*., "Delay-Tolerant Network Architecture," RFC 4838, April 2007.

[6] K. Scott and S. Burleigh, "Bundle Protocol Specification," RFC 5050, November 2007.

[7] L. Wood, J. McKim, W. M. Eddy, W. Ivancic and C. Jackson, "Using Saratoga with a Bundle Agent as a Convergence Layer for Delay-Tolerant Networking," work in progress as an internet-draft, draft-wood-dtnrg-saratoga, October 2008.

[8] L. Wood, W. Ivancic, W. M. Eddy, D. Stewart, J. Northam, C. Jackson and A. da Silva Curiel, "Use of the Delay-Tolerant Networking Bundle Protocol from space," paper IAC-08-B2.3.10, 59th International Astronautical Congress, Glasgow, September 2008.

[9] W. M. Eddy, L. Wood and W. Ivancic, "Checksum Ciphersuites for the Bundle Protocol," work in progress as an internet-draft, draft-irtf-dtnrg-bundle-checksum, October 2008.